\documentclass[showpacs,aps,prd,twocolumn]{revtex4}
\usepackage[latin1]{inputenc}
\usepackage{pstricks}
\usepackage{amsmath}
\usepackage{float}
\usepackage{color}
\usepackage{bm}
\usepackage{amssymb}
\usepackage[dvips]{graphicx,psfrag}

\newcommand{\be}{\begin{equation}}
\newcommand{\ee}{\end{equation}}
\newcommand{\beq}{\begin{eqnarray}}
\newcommand{\eeq}{\end{eqnarray}}

\newcommand{\rd}{{\rm d}}

\begin{document}
 \title{Modified gravity {\it \`{a} la } Galileon: Late time cosmic acceleration and observational constraints}

\author{Amna Ali$^1$, Radouane Gannouji$^2$, M. Sami$^1$}

\affiliation{$^1$ Centre of Theoretical Physics, Jamia Millia Islamia, New Delhi-110025, India}
\affiliation{$^2$ IUCAA, Post Bag 4, Ganeshkhind, Pune 411 007, India}

\begin{abstract}
In this paper we examine the cosmological consequences of fourth order Galileon gravity.  We carry out  detailed investigations of the underlying dynamics 
and demonstrate the stability of one de Sitter phase. The  stable de Sitter phase contains a Galileon field $\pi$ which is an increasing function of time $(\dot{\pi}>0)$. Using the required suppression of the fifth force, supernovae,  BAO, and CMB data,  we constrain 
parameters of the model. We find that the $\pi$ matter coupling parameter $\beta$ is constrained to small numerical values such that $\beta<0.02$. We also show that the parameters of the third and fourth order in the action $(c_3,c_4)$ are not independent and with reasonable assumptions,  we obtain constraints on them.  We investigate the growth history of the model and find that the sub-horizon approximation is not allowed for this model.  We demonstrate strong scale dependence of linear perturbations in the fourth order
Galileon gravity.

\end{abstract}

\maketitle

%%%%%%%%%%%%%%
\section{Introduction}
According to the dark energy paradigm, the phenomenon of late time cosmic acceleration \cite{DarkEnergy} can be understood  by assuming that an exotic relativistic fluid with large negative pressure fills the whole Universe \cite{DE}.  The simplest example of such a homogeneous fluid is provided by cosmological constant {$\Lambda$} which is automatically  present
in the Einstein equations by virtue of Bianchi identities. The dark energy model based upon the cosmological constant {\it \`a la} $\Lambda CDM$ is consistent with all the cosmological observations at present.  However,   the field theoretic understanding of $\Lambda$  is far from being satisfactory and its small numerical value leads to the well-known {\it coincidence} and
{\it fine tuning} problems.  A variety of scalar fields  as candidates of dark energy were then investigated in the hope of addressing the said problems. The scalar field models with
generic features might alleviate the fine tuning and coincidence problems leaving the dark energy metamorphosis as a challenge for future observations in cosmology.

There is an alternative view which advocates the necessity for the paradigm shift according to which late time cosmic acceleration is an artifact of large scale modification of 
gravity rather than the consequence of dark energy. On theoretical grounds it is plausible that gravity suffers modifications at large scales where it is never tested directly. 
We know that gravity gets corrected quantum mechanically at small scales which is beyond the observational reach at present.  As for the large scale modification,  it should 
give rise to late time acceleration, be distinguishable from $\Lambda$ and at the same time be consistent with local gravity constraints.  The last requirement is nontrivial
as Einstein gravity agrees with solar physics and the equivalence principal with a high accuracy.

There are broadly two ways used to evade the local gravity constraints namely the chameleon mechanism \cite{Chameleon} and Vainstein screening \cite{Vainshtein}. The first method is widely used in 
$f(R)$ theories of gravity \cite{fR} with a disappearing cosmological constant. In this scenario, the mass of the scalar degree of freedom dubbed scalaron
present in the theory  becomes large thereby hiding the scalaron locally.  
%%%%

In generic models
of $f(R)$ gravity \cite{fR2}, the chameleon mechanism allows us to satisfy
the local gravity constraints but at the same time makes these models
vulnerable to the problem of a curvature singularity whose resolution requires the
fine tuning worse than the one encountered in $\Lambda CDM$ model.
The problem can be alleviated by invoking an $R^2$ correction but the
scenario becomes problematic if extended to the early universe.
%%%
 The later puts  an stringent  constraint  on a viable large scale modification
of gravity within the framework of $f(R)$ theories.  

The scalar degrees of freedom also  naturally arise in the  four dimensional effective theories. They couple to matter source
and might give rise to fifth force effects. In this case, the local gravity constraints can be evaded using the Vainstein screening mechanism which is implemented  by using the nonlinear self interaction
of the scalar field. Nonlinearity becomes important in the vicinity of dense objects allowing the scalar degrees to decouple  from the matter source.   In DGP \cite{DGP}, the scalar degree
of freedom appears in the form of brane bending mode with the required nonlinear derivative interactions of the simplest type which is invariant under the shift symmetry in the
flat space time. The equation of motion for the scalar field dubbed Galileon is necessarily of second order. The general structure of higher order Lagrangian of the Galileon field
was obtained in \cite{Galileon}. Similar to Lovelock gravity, the Galileon gravity provides a consistent 
modification of GR leaving the local physics intact.  The DGP model which includes the lowest order
Galileon Lagrangian in its decoupling limit suffers from the problem of instabilities \cite{Inst}.
The Galileon modified gravity in its general setting can give rise to late time acceleration and 
is free from negative energy instabilities.  The model has the well-posed Cauchy problem
and is safe from paradoxes related to micro-causality \cite{Gannouji:2010au}.

In this paper we investigate the cosmological dynamics based upon Galileon gravity, set up the autonomous system and discuss the existence and stability of fixed points.
We especially  focus on the self-accelerating solution
and explore the observational constraints on the model parameters using supernovae, BAO and CMB data.  We also study  metric perturbations and investigate the growth
history of the model.

%%%%%%%%%%%%%%

%%%%%%%%%%%%%%
\section{Background}
%%%%%%%%%%%%%%
In Galileon  theories, the large scale modification  of gravity arises due to the nonlinear derivative self interaction of a scalar field $\pi$ dubbed the Galileon field,
which couples with matter and metric. In what follows, we shall consider the Galileon action of the form,

\be 
\mathcal{S}=\int{\textit{\rd}^4 x}\sqrt{-g} \left(\frac{R}{2}+c_\textit{i}~L^{(\textit{i})}\right)+\mathcal{S}_m[\psi_m,e^{2\beta\pi}g_{\mu\nu}]
\label{action}
\ee

where $\{c_i\}$ are constants and the $L_i^{'s}$ are given by

\beq
L^{(1)}&=&\pi\\
L^{(2)}&=&-\frac{1}{2}(\nabla\pi)^2 \equiv -\frac{1}{2} \pi_{;\mu}\pi^{;\mu}\\
L^{(3)}&=&-\frac{1}{2} (\nabla\pi)^2 \Box\pi\\
L^{(4)}&=&-\frac{1}{2}
(\nabla\pi)^2\left[(\Box\pi)^2-\pi_{;\mu\nu}\pi^{;\mu\nu}+\pi^{;\mu}\pi^{;\mu}G_{\mu\nu}\right]\nonumber \\
&&+(\Box\pi)\pi_{;\mu}\pi_{;\nu}\pi^{;\mu\nu}-\pi_{;\mu}\pi^{;\mu\nu}\pi_{;\nu\rho}\pi^{;\rho}
\eeq

The fourth order Galileon theory leads to the following evolution equations in the FRW background \cite{Gannouji:2010au},

\be
\label{eq:Fried1}
3H^2=\rho_m+\rho_r+\frac{c_2}{2}\dot{\pi}^2-3c_3 H\dot{\pi}^3+\frac{45}{2}c_4 H^2\dot{\pi}^4
\ee

\beq
\label{eq:Fried2}
2\dot{H}+3H^2&=&-\frac{1}{3}\rho_r-\frac{c_2}{2}\dot{\pi}^2-c_3\dot{\pi}^2\ddot{\pi}\nonumber \\
&&+\frac{3}{2}c_4\dot{\pi}^3\left(3H^2\dot{\pi}+2\dot{H}\dot{\pi}+8H\ddot{\pi}\right)
\eeq

\beq
\label{eq:KG}
\beta\rho_m&=&-c_2\left(3H\dot{\pi}+\ddot{\pi}\right)+3c_3\dot{\pi}\left(3H^2\dot{\pi}+\dot{H}\dot{\pi}
+2H\ddot{\pi}\right)\nonumber \\
&&-18c_4H\dot{\pi}^2\left(3H^2\dot{\pi}+2\dot{H}\dot{\pi}+3H\ddot{\pi}\right),
\eeq

where $H=\dot{a}/a$ is the Hubble function, $\rho_m$ and $\rho_r$ are the density of matter and radiation respectively
 
It was found \cite{Gannouji:2010au} that the model has a self-accelerating solution iff

\beq
&c_3^2-8c_2c_4>0,\\
&A_+>0~~\text{or}~~A_->0
\eeq

with $A_\pm=\frac{c_3^2-12 c_2c_4\pm c_3\sqrt{c_3^2-8c_2c_4}}{c_4}$.

We have, therefore, two de Sitter solutions for this model, namely,  the positive branch ($A_+$) and the negative branch ($A_-$).\\

In Ref. \cite{Gannouji:2010au} , various conditions of stability of the theory were derived.  It was  shown that   positive values for the parameters $(c_2,c_4,\beta)$ and $c_3>\sqrt{8c_2c_4}$ can give rise to viable evolution.\\

It is straightforward to show that $A_-<0$.  As for the  $A_-$, it  is a decreasing function of $c_3$ and therefore the highest value of $A_-$ is achieved when $c_3=\sqrt{8 c_2 c_4}$. We found $A_-^{\text{max}}=-4c_2<0$.\\

If we consider the conditions of the stability of the theory, then the negative branch is ruled out as it does not have a de Sitter phase thereby leaving 
us with  one self-accelerating solution in  the positive branch.
Bearing this in mind,  in the discussion to follow, we shall only consider  the positive branch of the theory and thus redefine $A=A_+$.\\

%%%%%%%%%%%%%%
\section{Autonomous system}
%%%%%%%%%%%%%%

For the sake of convenience, we use the system of units such that $\pi$ is dimensionless and so is also true for  $c_2$ .   We then define two new dimensionless parameters $\hat{c}_3=c_3 H_{dS}^{-2}$, $\hat{c}_4=c_4 H_{dS}^{-4}$ where $H_{dS}$ is the Hubble function during the de Sitter era.
It can easily be noticed that the parameters $(c_2,\hat{c}_3,\hat{c}_4)$ are not independent. In fact if we consider the de Sitter solution derived in Ref. \cite{Gannouji:2010au},
we find that,

\beq
H_{dS}\dot{\pi}_{dS}&=&\frac{c_3+\sqrt{c_3^2-8c_2 c_4}}{12 c_4}\\
48 H_{dS}^2&=&\dot{\pi}_{dS}^2 A,
\eeq

It is then straightforward to find the relation between $(c_2,\hat{c}_3,\hat{c}_4)$,

\be
\frac{48}{A}=\frac{\hat{c}_3+\sqrt{\hat{c}_3^2-8c_2 \hat{c}_4}}{12 \hat{c}_4}.
\ee

It is therefore obvious that the parameters are not independent if we normalized them by $H_{dS}$. On the other hand we can define a normalization of these parameters at any redshift, in particular we can do it today by

\beq
\label{definition1}
\tilde{c}_3&=&c_3 H_{0}^{-2}=\left(\frac{H_{dS}}{H_0}\right)^2\hat{c}_3\\
\label{definition2}
\tilde{c}_4&=&c_4 H_{0}^{-4}=\left(\frac{H_{dS}}{H_0}\right)^4\hat{c}_4
\eeq 

The factor $H_{dS}/H_0$ depends on the evolution of the model but for an evolution close to $\Lambda$CDM, we have

\beq
\tilde{c}_3&\simeq&\Omega_\Lambda\hat{c}_3\\
\tilde{c}_4&\simeq&\Omega_\Lambda^2\hat{c}_4
\eeq 

For  physical interpretation we will work with $(\tilde{c}_3,\tilde{c}_4)$  given by expressions  (\ref{definition1},\ref{definition2}).  In this case, it is reasonable to assume  that the parameters $(\tilde{c}_3,\tilde{c}_4)$ are of  order one. In fact it was showed in \cite{Gannouji:2010au} that the Galileon force is suppressed for scales smaller than $r_\star$ where $r_\star^3=\sqrt{\frac{\tilde{c}_4}{2\beta}}\frac{r_s}{H_0^2}$ and $r_s$ is the Schwarzschild radius of the source. If we consider $\tilde{c}_4$ and $\beta$ of the same order, we have  

\be
r_\star(\text{Earth})=1.7~\text{pc}
\ee
\be
r_\star(\text{Sun})=120~\text{pc}
\ee
\be
r_\star(\text{Milky Way})=1.2~\text{Mpc}
\ee

which are the same scales as in the DGP model.\\

The evolution equations  (\ref{eq:Fried1},\ref{eq:Fried2},\ref{eq:KG}) can easily be cast in the autonomous form.\\ 
Let $x=\dot{\pi}/H$ and $y=\dot{\pi} H/H_0^2$ with $H_0$ as  the Hubble constant today.
The evolution equations  acquire the form

\beq
\label{eq:autonomous1}
x'&=&\frac{\ddot{\pi}}{H_0^2}\frac{x}{y}-\frac{H'}{H}x\\
\label{eq:autonomous2}
y'&=&\frac{\ddot{\pi}}{H_0^2}+\frac{H'}{H}y\\
\label{eq:autonomous3}
\Omega_r'&=&-2\Omega_r\left(2+\frac{H'}{H}\right)
\eeq

where a prime represents a derivative with respect to $N =\ln ~a$ and

\begin{widetext}
\beq
\frac{\ddot{\pi}}{H_0^2}&=&\frac{3 y}{2 x}\frac{c_2 x \left(4+x^2 y (\tilde{c}_3-18 \tilde{c}_4 y)\right)
-\left(2-3 \tilde{c}_4 x^2 y^2\right) (3 \tilde{c}_3 x y-2 \beta  \Omega_m)+4 x y (\tilde{c}_3-12 \tilde{c}_4 y) \Omega_r}{54 \tilde{c}_4 x^2 y^3 (\tilde{c}_3-5 \tilde{c}_4 y)
+c_2 \left(-2+3 \tilde{c}_4x^2 y^2\right)-3 y \left(36 \tilde{c}_4 y+\tilde{c}_3 \left(-4+\tilde{c}_3 x^2 y\right)\right)}\\
\frac{H'}{H}&=&\frac{A_{x,y}}{108 \tilde{c}_4 x^2 y^3 (\tilde{c}_3-5 \tilde{c}_4 y)+c_2 \left(-4+6 \tilde{c}_4 x^2 y^2\right)-6 y \left(36 \tilde{c}_4
   y+\tilde{c}_3 \left(-4+\tilde{c}_3 x^2 y\right)\right)}\\
\Omega_m&=&1-\Omega_r-\frac{1}{6} x^2 \left(c_2-6 \tilde{c}_3 y+45 \tilde{c}_4 y^2\right)
\eeq
\end{widetext}

with

\beq
A_{x,y}&=&c_2^2 x^2+c_2 \left(6+3 x^2 y (-4 \tilde{c}_3+39 \tilde{c}_4 y)+4 \Omega_r\right)\nonumber\\
&&+18 \tilde{c}_3^2 x^2 y^2-6 \tilde{c}_3 y\left(6+45\tilde{c}_4 x^2 y^2+x \beta  \Omega_m+4 \Omega_r\right)\nonumber \\
&&-18 \tilde{c}_4 y^2 \left(18+45 \tilde{c}_4 x^2 y^2+4 x \beta  \Omega_m+12
   \Omega_r\right)
\eeq

Given the definition of the two variables $(x,y)$, it is enough to consider the phase space subject to the condition,  $sign(x)=sign(y)$.\\

This autonomous system has a saddle point, which corresponds to the radiation era:
\be
P_{r,1}:(x,y,\Omega_r)=(0,0,1)
% P_{r,2}&:&(x,y,\Omega_r)=(0,\frac{\tilde{c}_3\pm\sqrt{\tilde{c}_3^2-8c_2\tilde{c}_4}}{12 \tilde{c}_4},1)
\ee

The fixed point which corresponds to the matter-dominated epoch is 

\be
P_m:(x,y,\Omega_r)=(0,\infty,0),
% ~~~~\text{with}~~~1/y^2\ll x\ll 1/y,
\ee

whereas the two de Sitter points are given by

\be
P_{dS}:(x,y,\Omega_r)=(\pm\sqrt{\frac{48}{A}},\frac{\tilde{c}_3+\sqrt{\tilde{c}_3^2-8c_2\tilde{c}_4}}{12\tilde{c}_4},0)
\ee

We note that $y_{dS}$ is always positive because of the conditions of stability of the theory $(\tilde{c}_4>0,\tilde{c}_3>\sqrt{8c_2\tilde{c}_4})$.
 Hence, we shall  consider   only  those solutions which satisfy,  $x_{dS}>0$ $(sign(x_{dS})=sign(y_{dS}))$.  In this case, the system has only one de Sitter point  given by,

\be
P_{dS}:(x,y,\Omega_r)=(\sqrt{\frac{48}{A}},\frac{\tilde{c}_3+\sqrt{\tilde{c}_3^2-8c_2\tilde{c}_4}}{12\tilde{c}_4},0)
\ee

which is an attractor iff $A>16\beta^2/3$. In the case of physical interest with a small numerical value of  $\beta$ , we recover the condition ($A>0$) derived in \cite{Gannouji:2010au}. In this last case, the de Sitter point is always an attractor.\\

In addition to these three eras, various other critical points can be found. The relevant points are

\begin{widetext}
\be
P_1^\pm:(x,y,\Omega_r)=(\pm\sqrt{\frac{6}{c_2}},0,0),~~~~\Omega_m=0,~~~~w_{\rm eff}=1,~~~\text{saddle point}\nonumber
\ee
\beq
P_2:(x,y,\Omega_r)=(-\frac{2\beta}{c_2},0,0),~~~w_{\rm eff}=\frac{2\beta^2}{3c_2},~~\Omega_m=1-\frac{2\beta^2}{3c_2},~~\text{attractor if}~~\beta^2<\frac{c_2}{2}~~ \text{and a saddle point otherwise.}\nonumber
\eeq
\beq
P_3:(x,y,\Omega_r)=(\frac{3}{\beta},\sqrt{\frac{3c_2+2\beta^2}{27\tilde{c}_4}},0)&,&~~w_{\rm eff}=-1,~~\Omega_m=-4-9\frac{c_2}{\beta^2}+\frac{\tilde{c}_3}{\beta^2}\sqrt{\frac{9c_2+6\beta^2}{\tilde{c}_4}},~~\text{which is a saddle point} \nonumber\\
&& \text{or an attractor depending on the set of parameters $(c_2,\tilde{c}_3,\tilde{c}_4,\beta)$}.\nonumber
\eeq
\end{widetext}

\begin{figure}
\centerline{\includegraphics[scale=.6]{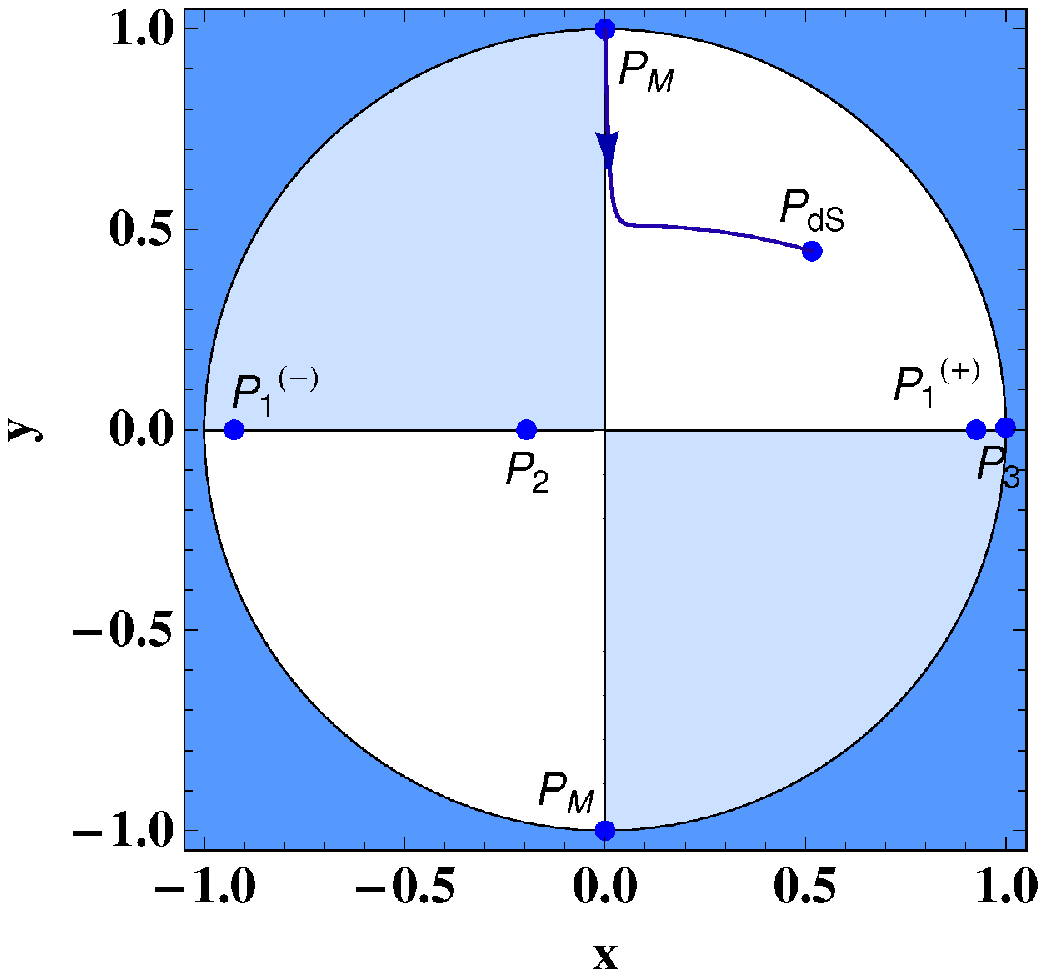}}
\centerline{\includegraphics[scale=.6]{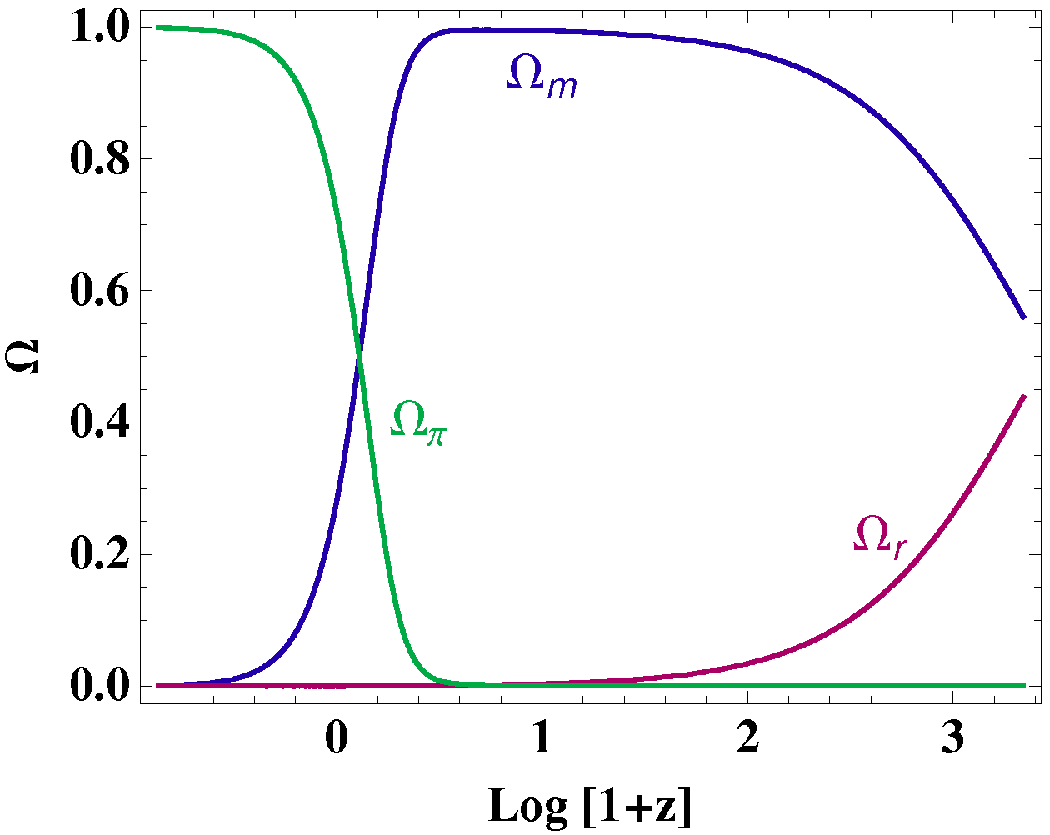}}
\caption{Top Panel: The projected phase space in the plane $(x,y)$ in Poincar\'e coordinates for $\beta=0.1$, $c_2=1$, $\tilde{c}_3=15$, $\tilde{c}_4=4$. The circles represent critical points. The initial conditions are chosen in the subspace $(x,y)\in \mathbb{R}_+^2$. Bottom Panel: The evolution of $\Omega$ as a function of $\log (1+z)$.}
\label{fig:dS}
\end{figure}

\begin{figure}
\centerline{\includegraphics[scale=.6]{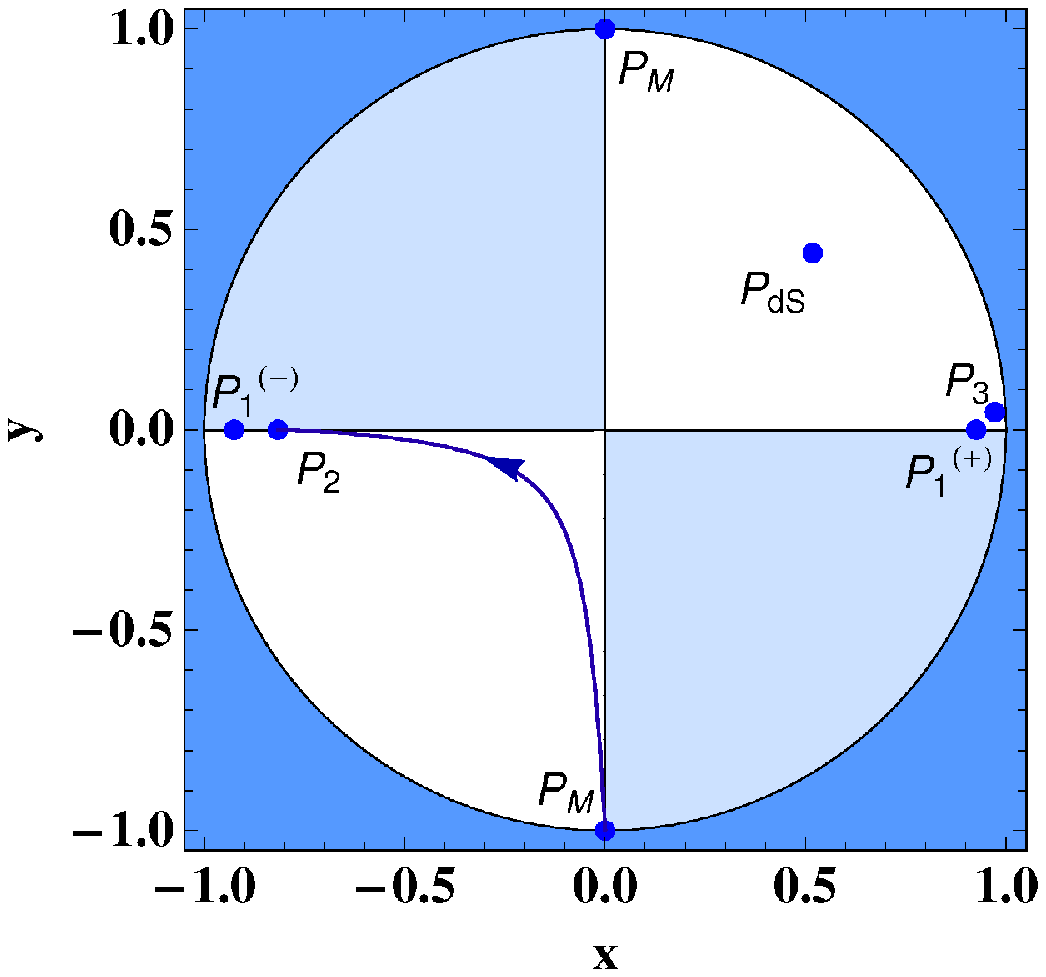}}
\centerline{\includegraphics[scale=.6]{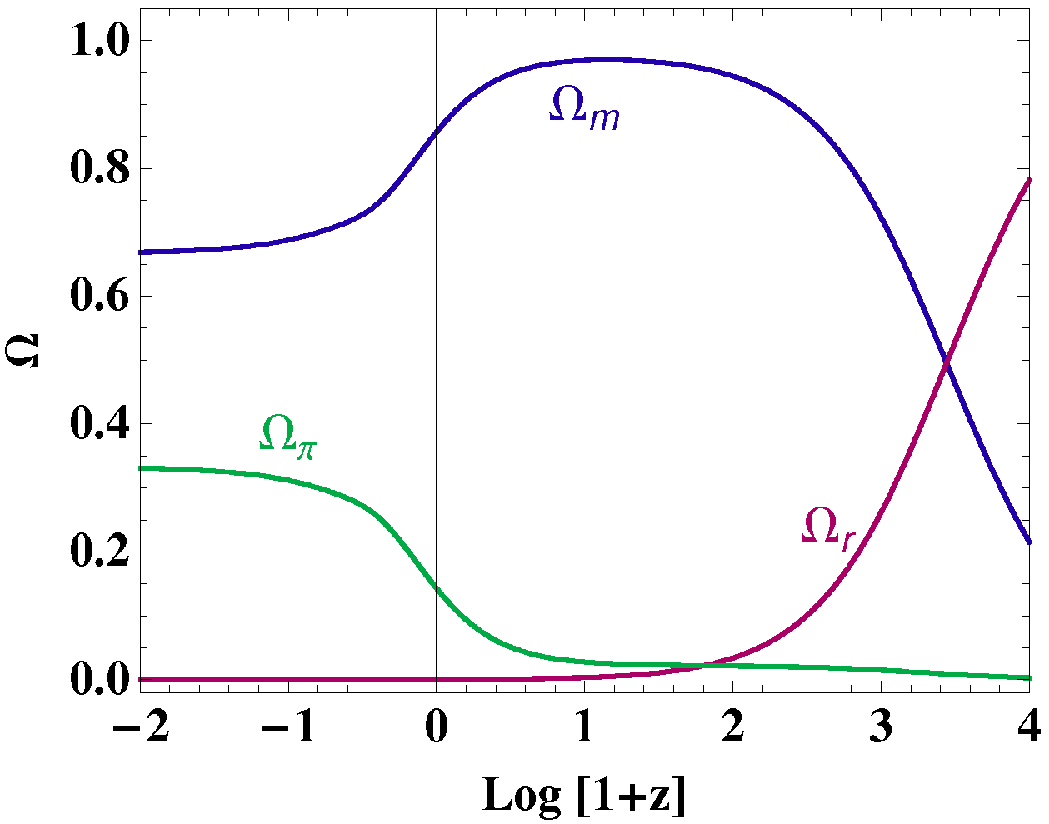}}
\caption{Top Panel: The projected phase space in the plane $(x,y)$ in Poincar\'e coordinates for $\beta=0.7$, $c_2=1$, $\tilde{c}_3=15$, $\tilde{c}_4=4$. The circles represent critical points. The initial conditions are chosen in the subspace $(x,y)\in \mathbb{R}_-^2$. Bottom Panel: The evolution of $\Omega$ as a function of $\log (1+z)$.}
\label{fig:P2}
\end{figure}

In Figs. \ref{fig:dS} and \ref{fig:P2}, we show the evolution of the autonomous system. We have a standard evolution in the first case where we chose positive initial conditions for $x$ and $y$. The system evolves along the axis $x=0$ during the matter phase before it is attracted by the de Sitter point along the line $y=y_{dS}$. We do not have a tracking solution when $\beta\neq 0$, but after the matter phase the coupling to matter $\beta$ is weak therefore we recover the tracking solution $y=y_{dS}$ derived in \cite{DeFelice:2010pv}.\\
In the second case we chose negative initial conditions, in this case there is no de Sitter point in the subspace considered, after a matter phase the model is attracted by $P_2$ (the attractor for these values of the parameters). We can see in the bottom plot of Fig. \ref{fig:P2} that in the future, $\Omega_m=1-2\frac{\beta^2}{3c_2}\simeq 2/3$  which corresponds to the attractor $P_2$.\\
It is  clear that we can not achieve a de Sitter phase if we consider initial conditions in the subspace $(x,y)<0$. Therefore, we have to consider positive initial conditions, which means $\dot{\pi}>0$.

\section{Observational constraints}

As demonstrated  in the previous section,  the parameters of the model are not free; we can consider $\tilde{c}_4$ as a function of $(\beta,c_2,\tilde{c}_3)$. Thus in this section, we shall impose the constraints on  the parameters $(\beta,c_2,\tilde{c}_3)$ only.\\
We constrain the parameters of the model by using supernovae, BAO and CMB datas. We used the compiled constitution set of $397$ type Ia supernovae \cite{Hicken:2009dk} for which the $\chi^2$ is defined as

\be
\chi^2_{SN1a}=\sum_{i}\frac{(\mu_{\rm th,i}-\mu_{\rm obs,i})^2}{\sigma_i^2}
\ee

with

\be
\mu_{th,i}=5\log~d_L(z_i)+\mu_0
\ee

where $\mu_0=25+\log~H_0^{-1}$ is marginalized \cite{DiPietro:2002cz,Lazkoz:2005sp} and $d_L$ is the luminosity distance.\\

We used the BAO distance ratio $D_v(z=0.35)/D_v(z=0.2)=1.736\pm 0.065$ \cite{Percival:2009xn}, where

\be
D_v(z)=\left[\frac{z}{H(z)}\left(\int_0^z\frac{\rd z'}{H(z')}\right)^2\right]^{1/3}
\ee

Finally we used the CMB shift parameter $R=1.725\pm 0.018$\cite{Komatsu:2010xm} , where

\be
R=H_0\sqrt{\Omega_{m,0}}\int_0^{z_{ls}}\frac{\rd z'}{H(z)}
\ee

\begin{figure}
\centerline{\includegraphics[scale=.7]{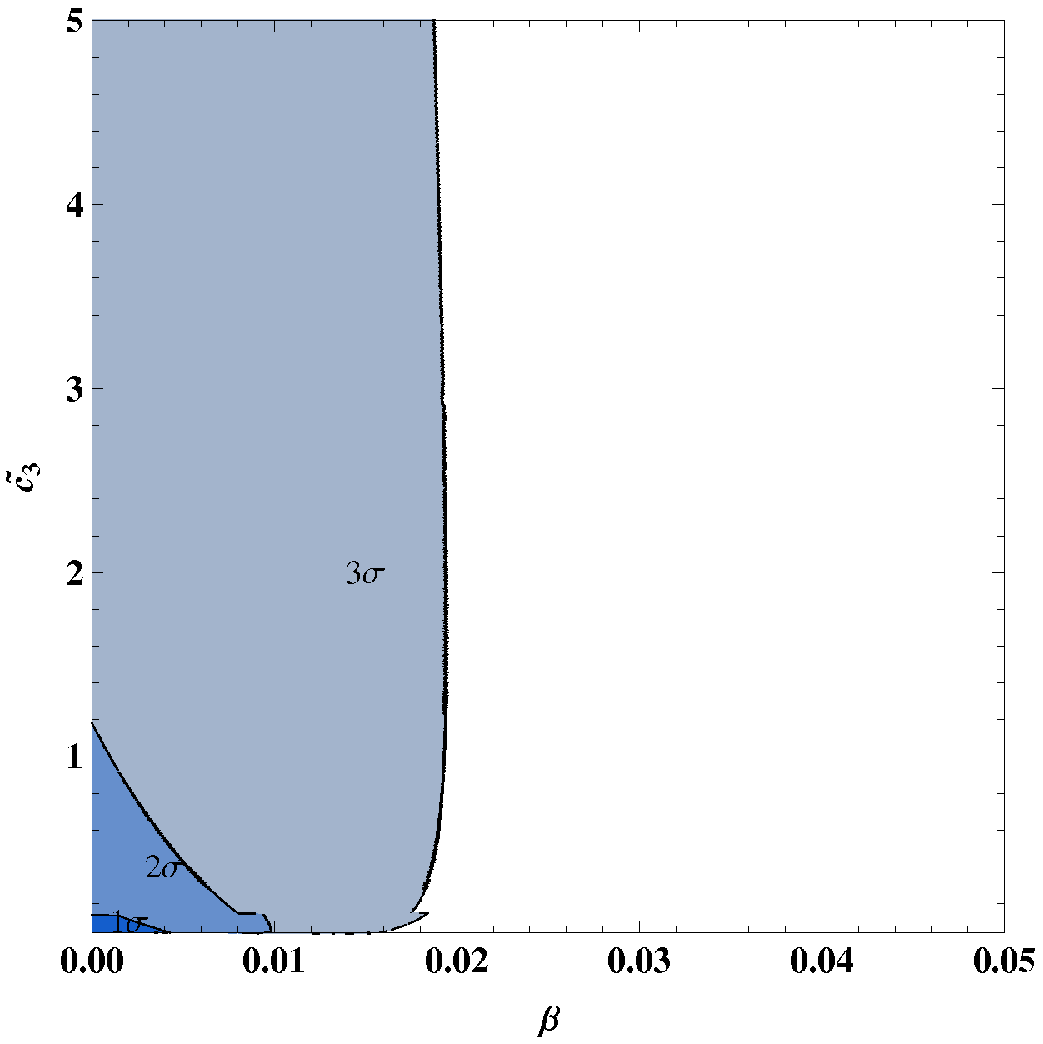}}
\centerline{\includegraphics[scale=.7]{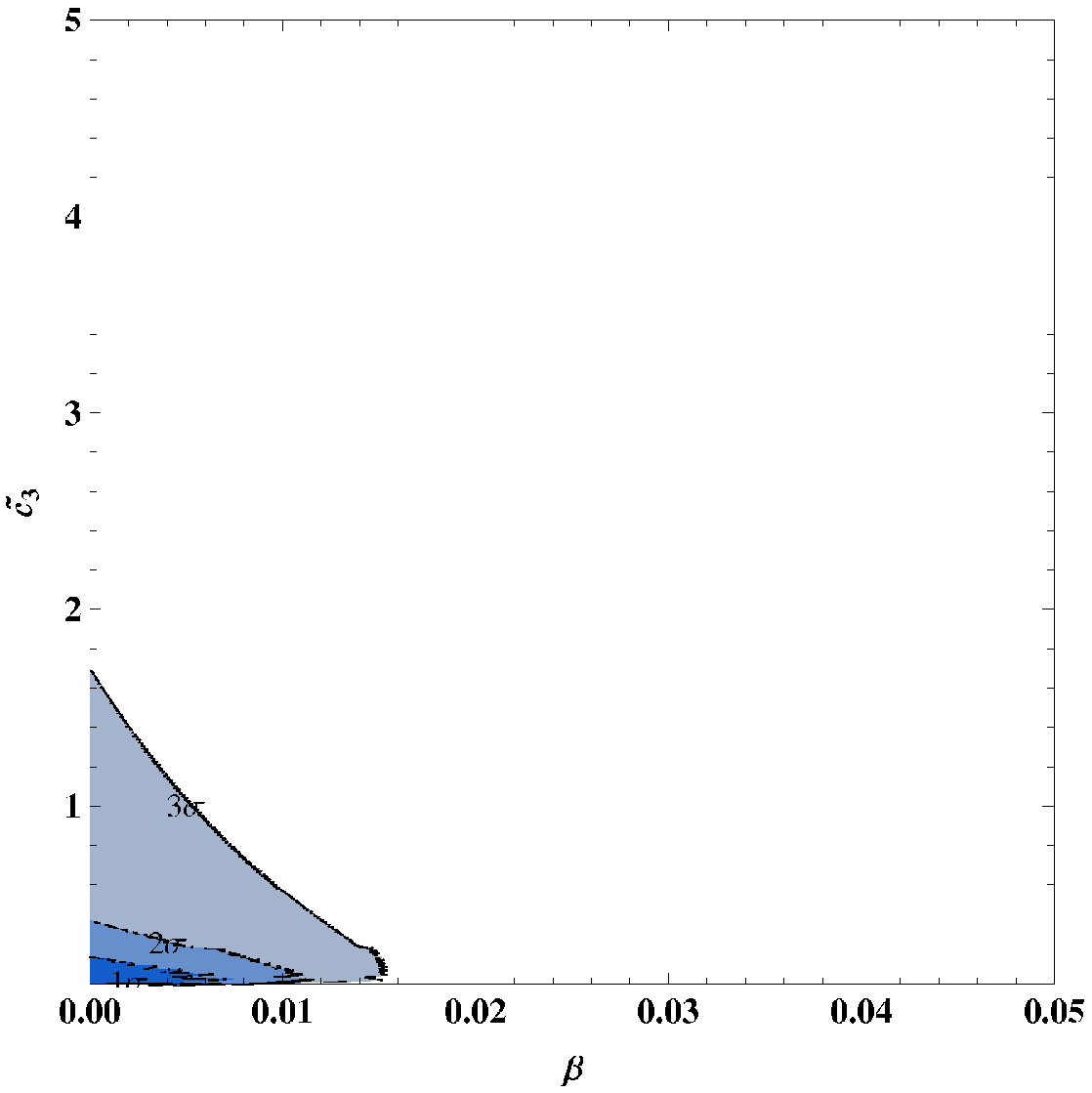}}
\centerline{\includegraphics[scale=.7]{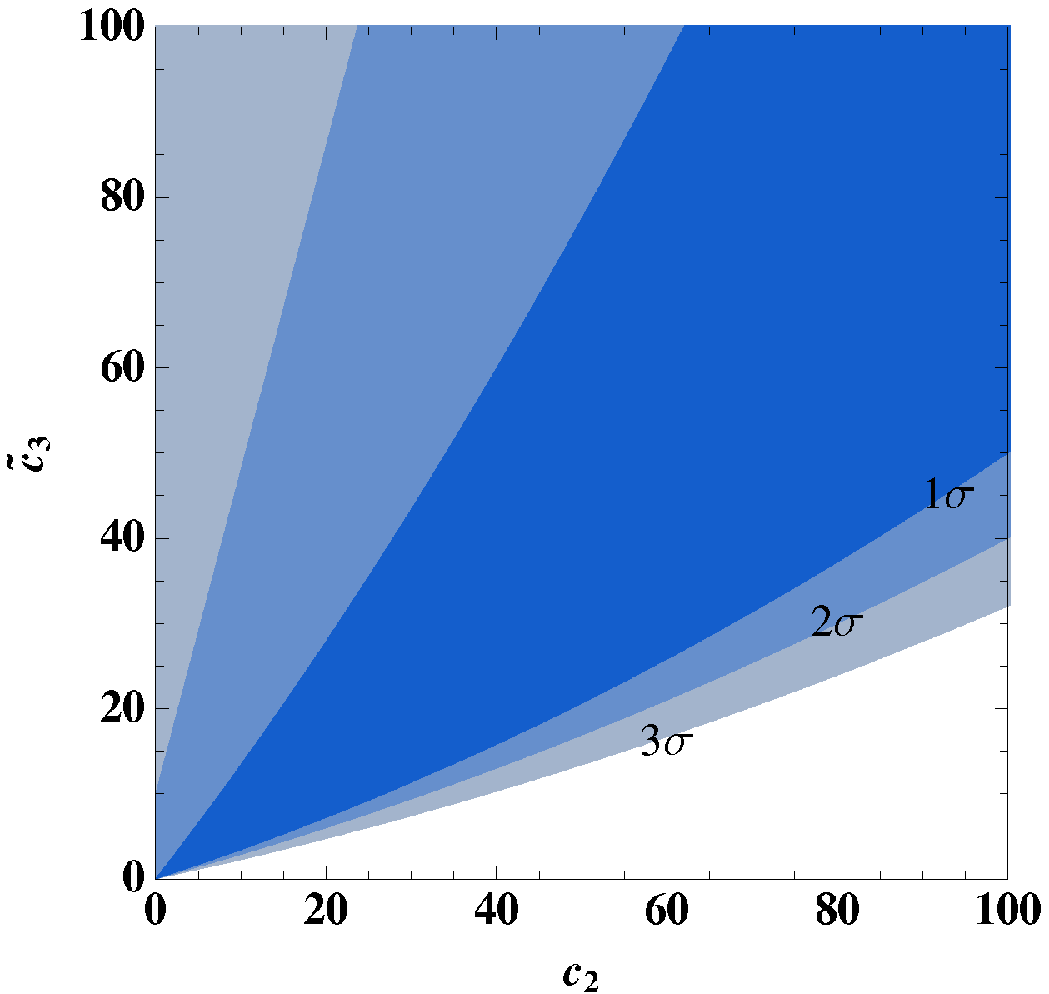}}
\caption{Top Panel: Contour plots at 1$\sigma$, 2$\sigma$ and 3$\sigma$ for $c_2=\beta$.\\Middle Panel: Contour plots at 1$\sigma$, 2$\sigma$ and 3$\sigma$ for $c_2=1$.\\Bottom Panel:Contour plots at 1$\sigma$, 2$\sigma$ and 3$\sigma$ for $\beta=0.01$.}
\label{fig:beta_c2}
\end{figure}

We observe (see Fig. (\ref{fig:beta_c2})) that when $c_2=\beta$ (motivated by a Galileon force of the order of the gravitational force at large scales \cite{Gannouji:2010au}), $\beta$ is constrained by the data to small values $\beta<0.02$, however the parameter $\tilde{c}_3$ is constrained at 2$\sigma$ at small values but unconstrained at 3$\sigma$.
For the model where we fixed $c_2=1$ we found that parameters are constrained to take small values even at 3$\sigma$.\\
In the bottom plot of the same figure, we chose a small value for the coupling $\beta=0.01$, we observe that for a fixed $c_2$ large values of $\tilde{c}_3$ are preferred.

% \begin{figure}
% \centerline{\includegraphics[scale=.7]{like.eps}}
% \caption{Likelihood plot for $\tilde{c}_3$ for $\beta=0.05$}
% \label{fig:like}
% \end{figure}

%%%%%%%%%%%%%%
\section{Perturbations}
%%%%%%%%%%%%%%

As mentioned earlier, the model is constrained differently by the data, when the assumption on $c_2$ is different. But we can always find a range of parameters which can fit the data, this is obvious because the model has a matter phase and a self-accelerating solution as for the $\Lambda$CDM. Therefore the large number of free parameters compared to the $\Lambda$CDM model cannot be constrained.\\
It is widely discussed in the literature that we can have a strong signature of modified gravity models by looking to the evolution of perturbations \cite{Pert}. We performed this analyze in the linear regime. We will consider, $k<0.1 h \rm Mpc^{-1}$, in this case we are at scales larger than $r_\star$, where the linear approximation is  allowed.\\
We will consider the perturbations in the comoving gauge. Therefore, we can write the equations of perturbations for $(\delta_m,\delta\pi)$ as the following:

\beq
\label{eq:pert1}
\ddot{\delta}_m+A_1\dot{\delta}_m+A_2\rho_m\delta_m+A_3\ddot{\delta\pi}+A_4\dot{\delta\pi}&+&A_5\delta\pi\nonumber\\
&=&0\\
\label{eq:pert2}
B_1\ddot{\delta}_m+B_2\dot{\delta}_m+B_3\rho_m\delta_m+B_4\ddot{\delta\pi}+B_5\dot{\delta\pi}&+&B_6\delta\pi\nonumber\\
&=&0
\eeq

The coefficients $\{A_i,B_i\}$ are given in the Appendix.

In the subhorizon approximation, we can approximate  Eq. (\ref{eq:pert2}) by

\be
\delta \pi=\frac{A_2B_1-B_3}{B_6-B_1A_5}\rho_m\delta_m
\ee

and this gives the equation for $\delta_m$ under this approximation

\be
\label{eq:deltam}
\ddot{\delta}_m+A_1\dot{\delta}_m-\frac{\rho_m}{2}G_{\rm eff}\delta_m=0
\ee

with

\be
G_{\rm eff}=-2\left(A_2+A_5\frac{A_2B_1-B_3}{B_6-B_1A_5}\right)
\ee

During the matter phase $x=\dot{\pi}/H \ll 1$ and $y=\dot{\pi}H/H_0^2 \gg 1$, therefore we have 

\beq
A_1&\simeq& 2H\,,\\
A_2&\simeq& -1/2\,,\\
B_6&\simeq& -y^2 \gg 1
\eeq

Then Eq. (\ref{eq:deltam}) can be approximated by

\be
\ddot{\delta}_m+2H\dot{\delta}_m-\frac{\rho_m}{2}\delta_m=0\,,
\ee 

We recover the equation of perturbations of standard general relativity during the matter phase. The evolution of $\delta_m$ for the Galileon model is equivalent to the evolution of matter perturbation in the standard model of cosmology. However after the matter phase, the approximation $x=\dot{\pi}/H \ll 1$ and $y=\dot{\pi}H/H_0^2 \gg 1$ is no longer valid and a deviation from the $\Lambda$CDM model appears.\\

\begin{figure}
\centerline{\includegraphics[scale=.7]{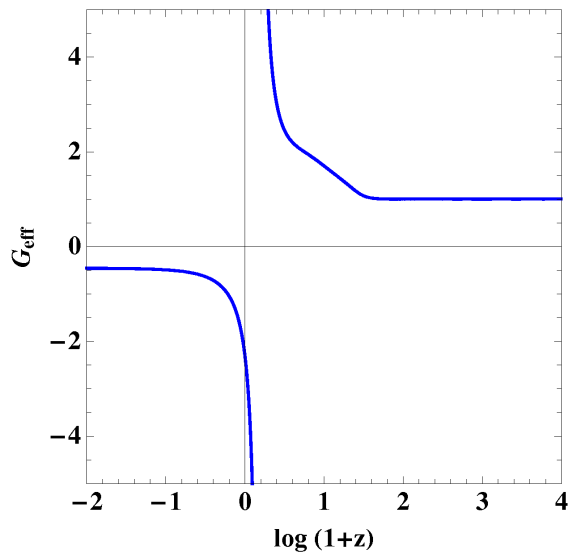}}
\caption{$G_{\rm eff}$ versus the redshift $\log(1+z)$ for $\beta=0.01$, $c_2=1$, $\tilde{c}_3=15$ and $\tilde{c}_4=4$.}
\label{fig:Geff}
\end{figure}
In Fig. \ref{fig:Geff} we can see that $G_{\rm eff}\simeq 1$ until $z\simeq 3.5$. But at small redshift $G_{\rm eff}$ diverges and the asymptotic value is negative.  It does not
necessarily imply the instability of the theory; this may be due to the subhorizon approximation that we did more than a problem of the theory. It was shown in Ref. \cite{Hwang:2009zj} that replacing a second-order differential equation by an algebraic equation (\ref{eq:pert2}) can be misleading.

Therefore we have to solve the complete system of coupled differential equations (\ref{eq:pert1},\ref{eq:pert2}).\\

\begin{figure}
\centerline{\includegraphics[scale=.7]{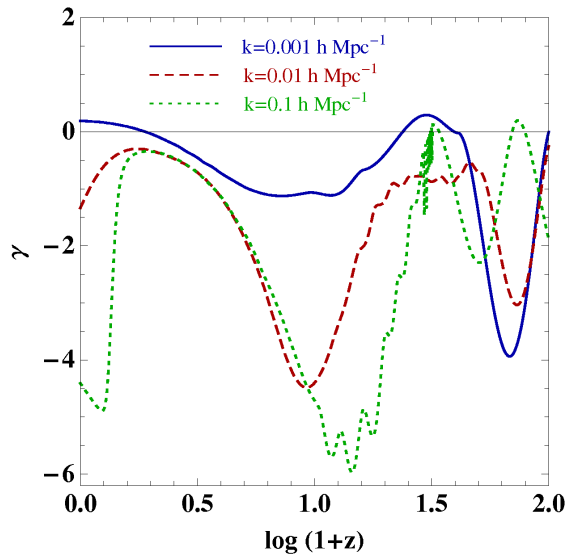}}
\caption{$\gamma$ versus $\log(1+z)$ for $\beta=0.01$, $c_2=1$, $\tilde{c}_3=15$ and $\tilde{c}_4=4$.}
\label{fig:gamma}
\end{figure}

Figure \ref{fig:gamma} shows $\gamma$  as a function of the redshift for the following parameters $(\beta,c_2,\tilde{c}_3,\tilde{c}_4)=(0.01,1,15,4)$ where $\gamma$ is defined by

\be
\gamma=\frac{\rm d \ln \delta_m}{\rm d \ln a}
\ee

For all the modes considered in the linear regime ($k<0.1 h \text{Mpc}^{-1}$), $\gamma$ has an oscillating mode during the matter phase; this oscillation becomes negligible for $z<1.5$. We remark that $\gamma$ is crucially different from its counterpart in models of dark energy within the framework of general relativity or modified gravity models as $f(R)$ or chameleon gravity. We have a strong dispersion of $\gamma_0$ with $\gamma_0(k=0.1~\text{h Mpc}^{-1})=-4.4$ and $\gamma_0(k=0.001~ \text{h Mpc}^{-1})=0.19$.

\begin{figure}
\centerline{\includegraphics[scale=.7]{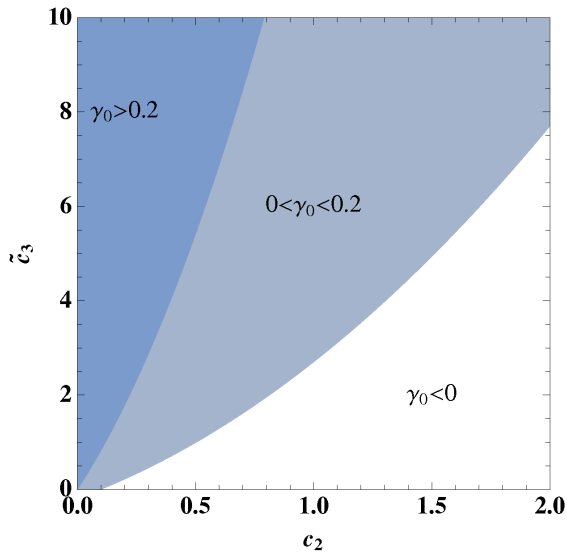}}
\caption{$\gamma_0\equiv\gamma(z=0)$ for different values of $(c_2,\tilde{c}_3)$ and $\beta=0.01$ for the scale $k=0.001~ \text{h Mpc}^{-1}$.}
\label{fig:gamma0}
\end{figure}

In the case considered in Fig. \ref{fig:gamma0} we showed the variation of $\gamma_0\equiv \gamma(z=0)$ for $0<c_2<2$ and $0<\tilde{c}_3<10$. We find that at small values of $c_2$, $\gamma_0$ is going larger $(\gamma_0>0.2)$ for the scale $k=0.001~ \text{h Mpc}^{-1}$. In the same range of $(c_2,\tilde{c}_3)$, $\gamma_0$ is shifted by $-4$ if we consider a smaller scale $k=0.1~ \text{h Mpc}^{-1}$.

\section{Conclusion}
In this paper, we have shown that  the Galileon modified gravity  has only one stable de Sitter  branch dubbed positive branch. We find that $\dot{\pi}$ should be positive for   the de Sitter phase to be reached. We have analyzed the stable branch by considering the constraints from Supernovae, BAO and CMB and found that the coupling parameter $\beta$ is constrained to small values, $\beta<0.02$.  The  parameters $(c_2,\tilde{c}_3,\tilde{c}_4)$ are not independent. Therefore we  restricted  our analysis to  two parameters namely, $(c_2,\tilde{c}_3)$ . We found that the contour plots are strongly dependent on the assumption. For $c_2=1$, motivated by a standard normalization of the kinetic term in the theory , we found that the model is strongly constrained even at 3$\sigma$. We also repeated our calculations for $c_2=\beta$ which is motivated by the requirement that the  Galileon force should be of the order of the gravitational force at large scales.  For this last case, the model is constrained at a small value of the parameters at 2$\sigma$ but $\tilde{c}_3$ and therefore, $\tilde{c}_4$ is unconstrained at 3$\sigma$.
Finally, we performed the analysis of  linear perturbations in the model under consideration.  We found that the subhorizon approximation can not be considered for the Galileon model, we have to consider the perturbations of the field too.  The growth of linear perturbations $\gamma$ turns out to be strongly dependent on the scales considered. At very large scales $\gamma_0 \simeq 0.25$ and this value is negative for smaller scales which can rule out the model.  It should be noted that the addition of the fifth term in the Galileon theory can produce different results as it contains an additional   free parameter $(c_5)$.  The inclusion of field  potential can also be considered to avoid the unwanted 
behavior in the linear regime of perturbations. In this  case, the model  can mimic features similar to the scenarios with a chameleon mechanism. In our opinion, it might be interesting to investigate this possibility.

\section{Acknowledgements}
RG thanks CTP, Jamia Millia Islamia for hospitality where a part of this work was carried out.

\begin{appendix}
\section{Perturbations}

\begin{widetext}
\beq
A_1&=&\frac{1}{\left(2-3 c_4 \dot{\pi }^4\right)^2}\left[8 H+\dot{\pi }^3 \left(-2 c_3+c_4 \left(-8 H \dot{\pi }+42 c_4 H \dot{\pi }^5-24 \ddot{\pi}-\dot{\pi }^4
   \left(c_3-36 c_4 \ddot{\pi}\right)\right)\right)\right]\\
A_2&=&-\frac{2+c_4 \dot{\pi }^4}{\left(2-3 c_4 \dot{\pi }^4\right)^2}\\
A_3&=&-\beta +3 \dot{\pi }^2 \frac{c_3-12 c_4 H \dot{\pi }}{-2+3 c_4 \dot{\pi }^4}\\
A_4&=&-3 H \beta +\frac{1}{\left(2-3 c_4 \dot{\pi }^4\right)^2}\left[-8 H \beta +80 c_4 H \beta  \dot{\pi }^4-150 c_4^2 H \beta  \dot{\pi }^8-4 \dot{\pi } \left(2 c_2+3 c_3
   \ddot{\pi}\right)+2 c_4 \dot{\pi }^5 \left(4 c_2+9 c_3 \ddot{\pi}\right)\right.\nonumber\\
&&\left.+9 c_4 H \dot{\pi }^6 \left(c_3-36 c_4 \ddot{\pi}\right) 
+18 H \dot{\pi }^2 \left(c_3+12 c_4 \ddot{\pi}\right)-4 \dot{\pi }^3 \left(18
   c_4 H^2+c_3 \beta -18 c_4 \dot{H}-6 c_4 \beta  \ddot{\pi}\right)\right.\nonumber\\
&&\left.-2 c_4 \dot{\pi }^7 \left(126 c_4 H^2-5
   c_3 \beta +54 c_4 \dot{H}+18 c_4 \beta  \ddot{\pi}\right)\right]\\
A_5&=&\frac{1}{\left(2-3c_4 \dot{\pi }^4\right)^2}
\left[
\frac{k^2}{a^2}\left(4 \beta -16 c_4 H \dot{\pi }^3-12 c_4 \beta  \dot{\pi }^4-24 c_4^2 H \dot{\pi }^7+9 c_4^2 \beta  \dot{\pi }^8+c_4
   \dot{\pi }^6 \left(c_3-36 c_4 \ddot{\pi}\right)+2 \dot{\pi }^2 \left(c_3+12 c_4 \ddot{\pi}\right)\right)\right.\nonumber\\
&&\left.-\beta  \left(24 H^2+48 c_4 H^2 \dot{\pi }^4+378 c_4^2 H^2 \dot{\pi }^8+3 \dot{H} \left(4-36 c_4 \dot{\pi }^4+45 c_4^2
   \dot{\pi }^8\right)+8 \dot{\pi }^2 \left(c_2+3 c_3 \ddot{\pi}\right)-4 c_4 \dot{\pi }^6 \left(2 c_2+9 c_3
   \ddot{\pi}\right)\right.\right.\nonumber\\
&&\left.\left.-12 c_4 H \dot{\pi }^7 \left(c_3-54 c_4 \ddot{\pi}\right)-24 H \dot{\pi }^3 \left(c_3+18
   c_4 \ddot{\pi}\right)\right)
\right]
-3 \beta  \dot{H}\\
B_1&=&-c_3 \dot{\pi }^2+12 c_4 H \dot{\pi }^3\\
B_2&=&
\dot{\pi } \left(c_2+6 c_4 \left(9 H^2+2 \dot{H}\right) \dot{\pi }^2-2 c_3 
\ddot{\pi}-6 H \dot{\pi } \left(c_3-6 c_4 \ddot{\pi}\right)\right)
-\frac{4 c_4 \dot{\pi }^2 \left(-2 H-c_3 \dot{\pi }^3+15 c_4 H \dot{\pi }^4\right) \left(H \dot{\pi }+3 \ddot{\pi}\right)}{-2+3 c_4 \dot{\pi }^4}
\\
B_3&=&-\beta +\frac{4 c_4 \dot{\pi }^2 \left(H \dot{\pi }+3 \ddot{\pi}\right)}{-2+3 c_4 \dot{\pi }^4}\\
B_4&=&-c_2+6 c_3 H \dot{\pi }+\left(c_3 \beta-54 c_4 H^2 \right) \dot{\pi }^2-12 c_4 H \beta  \dot{\pi }^3\\
B_5&=&\frac{4 c_4 \dot{\pi }^2 \left(-2 H \beta +c_2 \dot{\pi }-9 c_3 H\dot{\pi }^2+\left(90 c_4 H^2-c_3 \beta \right) \dot{\pi }^3
+15 c_4 H \beta  \dot{\pi }^4\right) \left(H \dot{\pi }+3 \ddot{\pi}\right)}{-2+3 c_4 \dot{\pi }^4}
-3 H \left(c_2-2 c_3 \ddot{\pi}\right)\nonumber\\
&&-3 H \dot{\pi }^2 \left(54 c_4 H^2-5c_3 \beta +36 c_4 \dot{H}+12 c_4 \beta  \ddot{\pi}\right)
+2 \dot{\pi } \left(9 c_3 H^2-c_2 \beta +3 c_3\dot{H}+\left(c_3 \beta-54 c_4 H^2 \right) \ddot{\pi}\right)\nonumber\\
&&-12 c_4 \beta  \left(12 H^2+\dot{H}\right) \dot{\pi }^3\\
B_6&=&
\frac{k^2/a^2}{-2+3 c_4 \dot{\pi }^4}\left[2 \left(26 c_4 H^2+c_3 \beta +12 c_4 \dot{H}\right) \dot{\pi }^2-24 c_4 H \beta  \dot{\pi }^3-3 c_4 \left(10
   c_4 H^2+c_3 \beta +12 c_4 \dot{H}\right) \dot{\pi }^6\right.\nonumber\\
&&\left.+36 c_4^2 H \beta  \dot{\pi }^7+2 \left(c_2-2 c_3 
\ddot{\pi}\right)-3 c_4 \dot{\pi }^4 \left(c_2+2 c_3 \ddot{\pi}\right)-8 H \dot{\pi } \left(c_3-6 c_4 
\ddot{\pi}\right)+8 c_4 H \dot{\pi }^5 \left(c_3+9 c_4 \ddot{\pi}\right)\right]\nonumber\\
&&+\frac{2 \beta}{-2+3 c_4 \dot{\pi }^4}\left[24 c_4 H \left(13 H^2+9 \dot{H}\right) \dot{\pi }^3-108 c_4^2 H \left(2 H^2+3 \dot{H}\right) \dot{\pi }^7+2 c_2
\ddot{\pi}+3 c_2 c_4 \dot{\pi }^4 \ddot{\pi}\right.\nonumber\\
&&\left.+6 H \dot{\pi } \left(c_2-4 c_3 \ddot{\pi}\right)-c_4 H
\dot{\pi }^5 \left(7 c_2+36 c_3 \ddot{\pi}\right)-12 \dot{\pi }^2 \left(c_3 \dot{H}+3 H^2 \left(c_3-8 c_4
\ddot{\pi}\right)\right)\right.\nonumber\\
&&\left.+6 c_4 \dot{\pi }^6 \left(3 c_3 \dot{H}+H^2 \left(5 c_3+54 c_4 \ddot{\pi}\right)\right)\right]
\eeq
\end{widetext}

\end{appendix}

\end{document}